\documentclass[10pt]{article}
\usepackage{multicol}
\usepackage{wrapfig}
\usepackage{authblk}
\usepackage[utf8]{inputenc}
\usepackage{geometry}
\usepackage{caption}
\usepackage{ulem}
\usepackage{amsmath}
\usepackage{graphicx}
\title{Electric Field Intensity Modulated Scattering \\as a Thin-Film Depth Probe}
\author[a,b*]{Peter J. Dudenas}
\author[b]{Adam Z. Weber}
\author[b]{Ahmet Kusoglu}

\affil[a]{Chemical and Biomolecular Engineering, University of California, Berkeley, CA 94720, USA}
\affil[b]{Energy Conversion Group, Lawrence Berkeley National Lab, CA 94720, USA}

\begin{document}
\date{}
\maketitle                        % DO NOT DELETE THIS LINE

% \begin{synopsis}
% A method of collecting and analysing grazing incidence x-ray scattering images versus incidence angle is developed and validated using two sets of polymer thin films.
% \end{synopsis}

\begin{abstract}
Grazing incidence x-ray scattering provides nanostructural information for thin film samples, but single images generally do not provide information on film thickness or the full complex index of refraction. Additionally, for thin films that possess stratification of scatterers vertically within a sample, it can be difficult to determine where those scatterers are positioned. We present an \textit{in-situ} method to extract film thickness, the index of refraction, and depth information using scattering images taken across a range of incident angles. The underlying theory is presented, and we validate the technique using two sets of polymer thin films. Finally, we discuss how it can be implemented as a general beamline procedure. This technique is applicable to any thin-film material and has potentially far-reaching impact by enabling depth-sensitive information \textit{in situ} at any grazing incidence-capable beamline.
\end{abstract}

     %-------------------------------------------------------------------------
     % The main body of the paper
     %-------------------------------------------------------------------------
     % Now enter the text of the document in multiple \section's, \subsection's
     % and \subsubsection's as required.
\begin{multicols}{2}
\section{Introduction}
\par Grazing incidence x-ray scattering (GIXS) has become a routine characterization method to determine the nanostructure morphology of thin films. Third generation synchrotron sources have the x-ray flux and detector technology necessary to provide high-throughput environments in which researchers can shoot their samples quickly and efficiently. Interpreting data is often the limiting step to completing GIXS studies, which is made more complicated due to dynamical scattering. This can make it necessary to model data within the Distorted-Wave Born Approximation (DWBA). Despite a number of programs available to model GIXS data (IsGISAXS\cite{RN1}, HiPGISAXS\cite{RN3}, BornAgain\cite{RN2}), modeling is often still a challenging and computationally expensive process. 
\par To model data properly, the appropriate form and structure factors need to be used, which may not be easily determined. Fresnel coefficients for each layer of the sample are also required, but are often variables in fitting experimental data. For samples with multiple layers this can quickly increase the number of fitting parameters. X-ray reflectivity (XRR) can be used to determine these optical parameters, but is most often done \textit{ex situ} because many beamlines do not have simultaneous capability for reflectivity and scattering. Electric field intensity (EFI)-modulated scattering provides a way to determine the optical constants and depth information of a sample \textit{in situ}. This is done by modeling the modulation in scattering intensity due to changes in the EFI and does not require knowledge of the form and structure factors.
\par In this paper, we summarize the relevant theory for grazing incidence scattering and optics, and then demonstrate how GIXS data taken as a function of incidence angle can be used to determine the index of refraction and film thickness of a sample, as well as determine where specific scatterers exist within a film. Real-beam effects such as energy resolution and angular divergence are considered. Finally, we outline how this procedure could be automated at the beamline to help researchers perform material exploration more intelligently.

\section{Theoretical Background}
\par A large body of literature exists covering theory of the DWBA scattering cross-section in grazing-incidence geometry for both x-ray and neutron scattering\cite{RN7,RN4,RN5,RN8,RN6,RN14}; we briefly outline past work before extending these results to EFI-modulated scattering. 
\par The total scattering cross-section for a rough film is shown below in Equation 1\cite{RN9}.
\begin{equation}
\frac{d\sigma}{d \Omega}=\left(\frac{d \sigma}{d \Omega}\right)_{Fresnel} e^{-Q_z^2 \sigma^2}+\left(\frac{d \sigma}{d \Omega}\right)_{diffuse}\
\end{equation}
The first term in Equation 1 is the specularly reflected x-ray beam, which is the Fresnel reflectivity multiplied by a Debye-Waller like factor that dampens intensity with increasing surface roughness, $\sigma$. This is what is measured in XRR experiments, but is often blocked in grazing incidence x-ray experiments to prevent the intense specular reflection from damaging the area detector. The second term in Equation 1 is the diffuse scattering cross-section, which includes scattering due to interfacial roughness, and scattering from any density fluctuations within the film. DWBA modifies the scattering cross-section from the Born Approximation (BA) to account for multiple scattering events and has been generalized to rough multilayers for specular and non-specular XRR\cite{holy1993x,holy1994} and diffuse scattering from buried nanostructures\cite{RN14}
\par The multilayer DWBA expression for diffuse scattering derived by Jiang et. al is 

\begin{multline}
         \left(\frac{d \sigma}{d \Omega}\right)_{diffuse} \approx \\ 
        r_e^2\left\langle\left |\sum_{j=1}^{N+1}\sum_{m=1}^4 \Delta\rho_j D_j^m F_j(\boldsymbol{q_{||}},q_z^m)S_j(\boldsymbol{q_{||}},q_z^m)\right |^2\right\rangle
\end{multline}
where $r_e$ is the classical electron radius, $\Delta\rho$ is the electron density contrast, $F(q_{||},q_z^m)$ is the form factor, $S(q_{||},q_z^m)$ is the structure factor, and $D_j^m$ and $q_z^m$ are defined as
$$D_j^1 = T_j^f T_j^i\qquad q_j^1 = k_{z,j}^f - k_{z,j}^i$$
$$ D_j^2 = R_j^f T_j^i\qquad q_j^2 = -k_{z,j}^f - k_{z,j}^i$$
$$ D_j^3 = T_j^f R_j^i\qquad q_j^3 = k_{z,j}^f + k_{z,j}^i $$
$$ D_j^4 = R_j^f R_j^i\qquad q_j^4 = -k_{z,j}^f + k_{z,j}^i. $$

$T_j^{i,f}$ and $R_j^{i,f}$ are the transmitted and reflected complex amplitudes of the incoming and final waves, respectively. The z-component of the wavevector for each of these transmitted and reflected waves is $\pm k_{z,j}^{i,f}$. 
\par Equation 2 takes in to account the local electric field intensity (EFI) varying across the depth of a scatterer, which can distort significantly the observed scattering pattern. EFI is defined as
\begin{equation}
	EFI = \left|T^i e^{-i k_{z}^i z} + R^i e^{ik^i_{z} z}\right|^2
\end{equation}

\par Equation 2 can be cast into a form that shows an explicit dependence on the EFI. To do this, we note that the $q^m_z$ dependence of the form and structure factors is just $e^{i q^m_z(z_j) z_j}$. Pulling this term out, part of Equation 2 can be refactored to show the initial and final wave eigenstates within the cross-section
\begin{multline}
    \sum_{m=1}^4 D_j^me^{i q_{z}^m (z_j) z_j} =\\
    \Big (T_j^iT_j^fe^{iq_z^1(z_j)z_j} + T_j^i R_j^f e^{i q_z^2(z_j)z_j}\\
    + T_j^f R_j^i e^{i q_z^3(z_j)z_j} + R_j^i R_j^f e^{i q_z^4(z_j)z_j} \Big)
    \end{multline}
\begin{multline}
   \sum_{m=1}^4 D_j^me^{i q_{z}^m (z_j) z_j} = \\
   \left (T_j^i e^{-i k_{z,j}^i z_j} + R_j^i e^{ik^i_{z,j} z_j}\right) \left ( T_j^f e^{i k_{z,j}^f z_j} + R_j^f e^{-ik^f_{z,j} z_j}\right ) 
\end{multline}
where the first and second grouping of terms are the z-components of the initial ($\psi^i_{j}$) and final ($\psi^f_{j}$) wave eigenstates, respectively. Plugging these back into Equation 2, we arrive at
\begin{multline}
		\left(\frac{d \sigma}{d \Omega}\right)_{diffuse} \approx \\
        r_e^2\left\langle\left |\sum_{j=1}^{N+1}\Delta\rho_j \psi^i_{z,j} \psi^f_{z,j} F_j(\boldsymbol{q_{||}})S_j(\boldsymbol{q_{||}})\right |^2\right\rangle.
\end{multline}
Continuing, if the scattering between layers is not coherent, then the sum can be taken outside the squared modulus and ensemble averaging. This leads to
\begin{multline}
		\left(\frac{d \sigma}{d \Omega}\right)_{diffuse} \approx \\
        r_e^2\sum_{j=1}^{N+1}|\Delta\rho_j|^2 |\psi^i_{z,j}|^2 |\psi^f_{z,j}|^2 \langle |F_j(\boldsymbol{q_{||}})|^2\rangle |S_j(\boldsymbol{q_{||}})|^2.
\end{multline}
Comparing Equations 3, 5, and 7, one can see that $|\psi_{z,j}^i|^2$ is the EFI within the layer j and the diffuse scattering of layer j is directly proportional to it. This provides the theoretical underpinning for EFI-modulated scattering. Incidence angle resolved (IAR) data is generated by collecting scattering images across a range of incidence angles. Intensity is sampled from a specific spot on the detector at each of these incidence angles to generate intensity versus angle, or IAR plots. This spot on the detector will be at a constant $k_{z}^f$ and $\boldsymbol{q}_{||}$, so the only varying component of the diffuse scattering cross section is the local EFI present wherever the scatterer resides that is contributing to the signal on the detector (i.e. I($\alpha_i$) $\propto$ EFI($\alpha_i)$). In this way, IAR data can be modeled using EFI calculations to provide depth information about the sample, including sample thickness, index of refraction, and where scatterers reside within a film. This modeling is simpler compared to traditional DWBA calculations because the form and structure factors do not need to be known; they are constant at fixed $k_{z}^f$ and $\boldsymbol{q}_{||}$. Finally, while this may be more complicated for systems where scattering is correlated between layers, modeling IAR data with EFI calculations may be a good approximation if the cross terms are small in Equation 6.
\par To calculate the EFI throughout a film, we use Parratt’s recursion\cite{RN15}, as laid out in M. Tolan’s monograph \textit{X-Ray Scattering from Soft-Matter Thin Films}\cite{RN17} (see SI for equations). Calculating the EFI through a film as a function of incidence angle leads to EFI maps as shown in Figure 1. 
\begin{center}
\includegraphics[width=\linewidth]{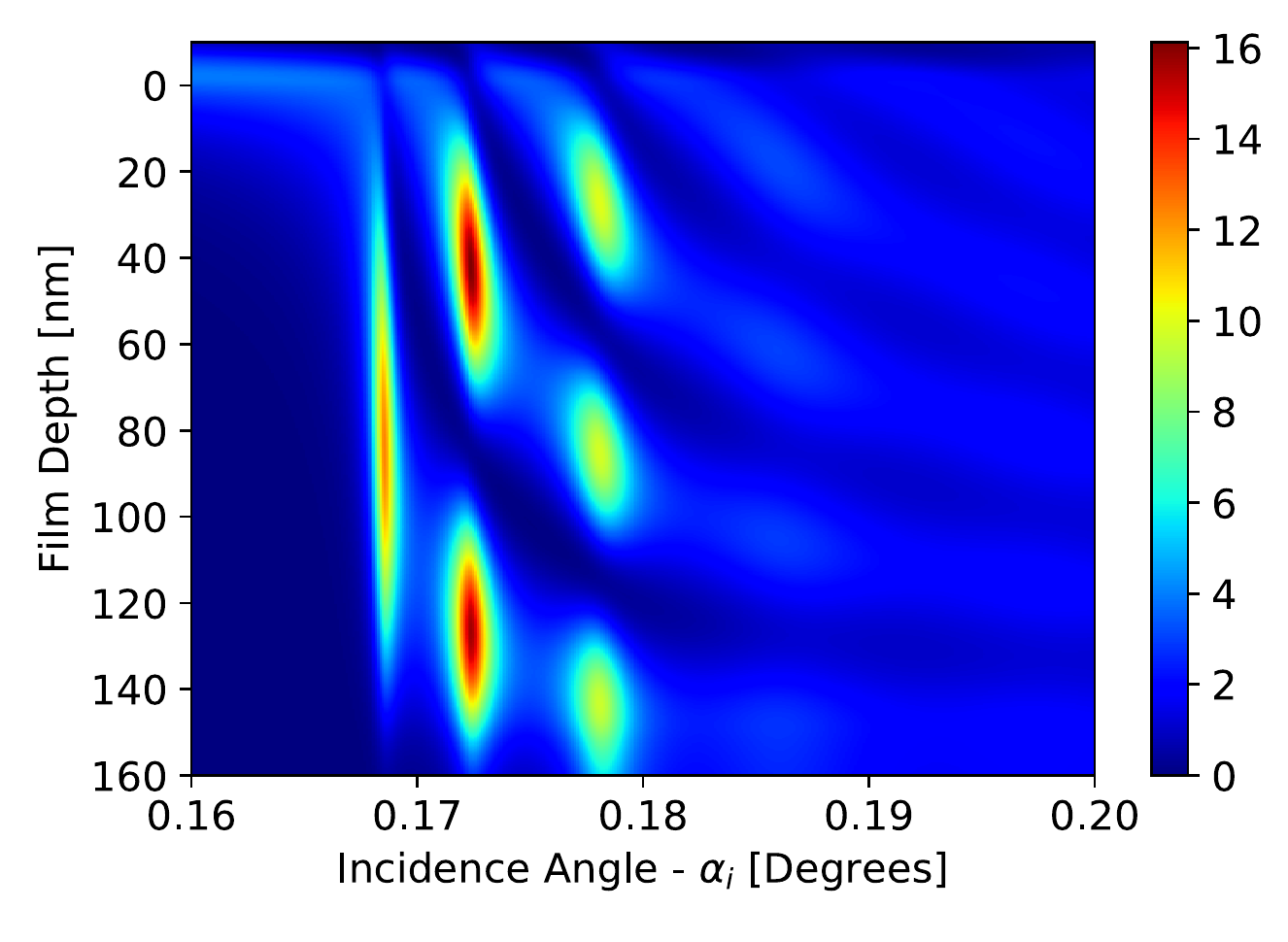}
\captionof{figure}{EFI map for a 150 nm Nafion thin film at 10 keV. Nafion is a perfluorinated sulfonic-acid ion-conducting polymer commonly used in fuel cells. The air-film and film-substrate interfaces are at 0 nm and 150 nm, respectively.}
\end{center}
\par At specific incidence angles, the incoming x-ray beam couples into the film, leading to transverse electric (TE) modes and large EFI enhancement at certain depths within the film. The angle- and depth-dependency of the EFI is what allows researchers to determine the z-position of specific atoms or scatterers within a sample. EFI-modulated fluorescence has been used to depth-probe surface chemistries\cite{bedzyk82,bedzyk89,bedzyk90} and GIXS data modeled with the multilayer DWBA has elucidated the positions of nanoparticles within polymer thin films \cite{RN10,RN11,RN12,RN13}. More recently, EFI principles were applied to reveal stratification of crystallite orientation in semi-conducting polymer thin films\cite{RN16}, based on modulation in diffraction peak intensities. Our work here applies these principles to develop a general depth-probe for thin-film systems using the specular rod signal in GIXS.

\section{Real-Beam Effects}
\par The EFI map in Figure 1 is calculated assuming a monochromatic plane wave. At a beamline, sources will have a finite angular divergence and energy resolution and may need to be accounted for depending on the magnitude of these values. These effects are considered using the energy resolution and angular divergence from Beamline 7.3.3 at the Advanced Light Source\cite{RN19}, where the experimental data was taken. 

\subsection{Energy Resolution}
\par 7.3.3 uses a pair of multilayer mirrors as their monochromator, where each mirror is 250 multilayers of alternating B$_4$C and Mo. The multilayer period is 2 nm ($\Gamma$=0.5) and the monochromator is operated at 10 keV ($\lambda$=1.24 \AA{}), with an incidence angle of 1.794 degrees (1st multilayer Bragg angle).  The FWHM is $\sim$100 eV and the full energy spectrum is shown in Figure S1. To calculate the effect of finite energy resolution, EFI maps at each energy are calculated and then summed, weighted by the energy spectrum probability distribution function. Correspondingly, the index of refraction for the film and substrate must also be adjusted for each energy, which are calculated here using the Center for X-Ray Optics (CXRO) online calculator\cite{RN18}. 
\par Figure 2a compares the average EFI of a polymer thin film on silicon for a monochromatic wave with the energy resolution for 7.3.3. The relatively large energy resolution for 7.3.3 reduces the modulation magnitude and slightly shifts the positions of TE modes. This will be taken in to account for further calculations below. For beamlines that use a Si $\langle 111\rangle$ monochromator (FWHM $\sim$2 eV), this effect should be negligible at hard x-ray energies. More broadly, the impact of energy resolution will depend on how quickly the index of refraction changes for a given material, which is a function of the chemical makeup and the energy range over which it is considered. It is expected that energy resolution will have a greater effect at lower energies where the index of refraction changes more quickly with energy.
\begin{center}
\includegraphics[width=0.85\linewidth]{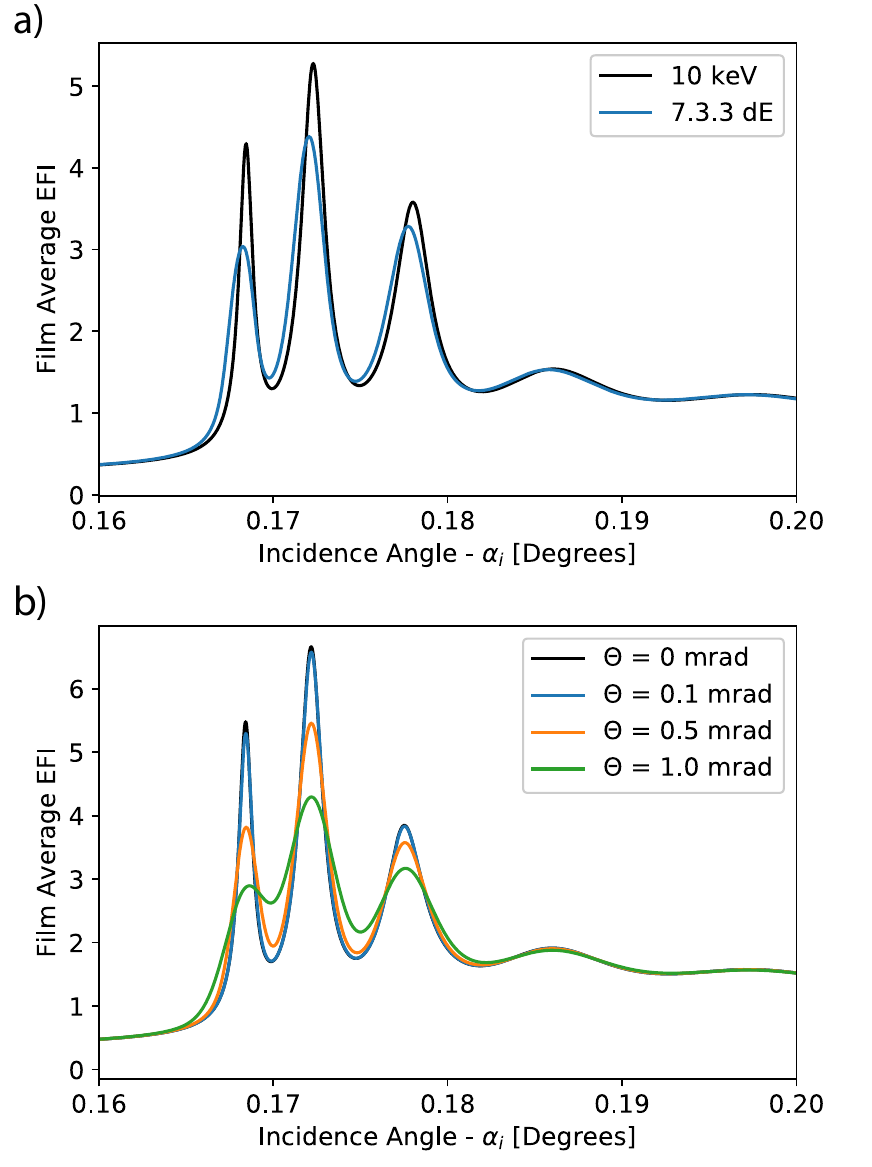}
\captionof{figure}{a) The average EFI for a 150 nm Nafion film calculated at 10 keV and the energy resolution of 7.3.3. b) Average EFI for the same Nafion film calculated for various values of angular divergence. 7.3.3's angular divergence is measured to be less than 0.1 mrad.}
\end{center}
\subsection{Angular Divergence}
\par The second effect that may need to be accounted for is angular divergence of the beam. At 7.3.3, this has been measured to be less than 0.1 mrad\cite{drizzle}. The divergence is approximated by convolving the EFI for a plane wave with a Gaussian distribution in incidence angle. Figure 2b shows the effect of multiple values of angular divergence. For 7.3.3's angular divergence, the effect on the EFI is negligible, and will be neglected for further calculations.

\section{Experimental Results\\ and Discussion}
\par We demonstrate EFI-modulated scattering first with a set of homogeneous polymer thin films of varying thickness to show how IAR data can be used to determine the index of refraction and thickness. Second, a set of bilayer samples is analyzed to show that multiple sources of scattering can be depth resolved within a film. The data collection and analysis for both sets of films is the same: Short-exposure GIXS images are taken at a large number of incidence angles and the same area from each image is binned into an intensity value to generate intensity versus incidence angle or IAR plots. In addition to sampling scattering peaks from a sample's internal structure, IAR data is also generated from the specular rod. This specular rod IAR data is what is fit using EFI calculations, due to its strong scattering signal. Figure S2 schematically shows the data collection and analysis procedures.
\par The specular rod signal is due to interfacial roughness which causes the incoming x-ray beam to be scattered/reflected with some small change in q$_x$, q$_y$;  this is proportional to the EFI at each of the interfaces present within a film. In the same way that knowledge of form and structure factors is not required, the specular rod signal can be broken down into the relative contributions from each interface without having to know the respective height-height correlation functions. Below the critical angle ($\alpha_c$), an evanescent wave penetrates only a few nanometers into the film and for layers thicker than this penetration depth, only the top interface contributes to the specular rod signal for $\alpha_i < \alpha_c$. An EFI map for a film is calculated given input parameters of thickness, and the real ($\delta$) and imaginary ($\beta$) parts of the index of refraction for each layer. The top surface from the calculated map is scaled to the specular rod IAR data below the critical angle and subtracted. The remaining intensity is from the EFI at the bottom interface for single layer films and from a combination of the EFI at the remaining interfaces in multilayer films. For very thin films, the evanescent wave will "see" the bottom layer below the critical angle, which may necessitate simultaneous fitting of the relative contributions.  Each of these procedures are incorporated into respective differential evolution fitting routines to minimize error between the data and model using the SciPy python package\cite{scipy}.

\subsection{Single Layer Films}
\par Thin films of Nafion (an ion-conducting polymer commonly used in fuel cells) were cast at 8 different thicknesses and sets of GIWAXS images were taken at incidence angles from 0.12$^{\circ}$-0.25$^{\circ}$.  Film thickness, $\delta$, $\beta$, and a scaling constant are the fitting parameters. Additionally, the sample alpha alignment procedure at 7.3.3 is not perfect and appears to shift the true incidence angle by up to 0.02$^{\circ}$. This offset is determined by externally iterating through offset values to find a minimum in the fitting error.
\begin{center}
\includegraphics[width=0.85\linewidth]{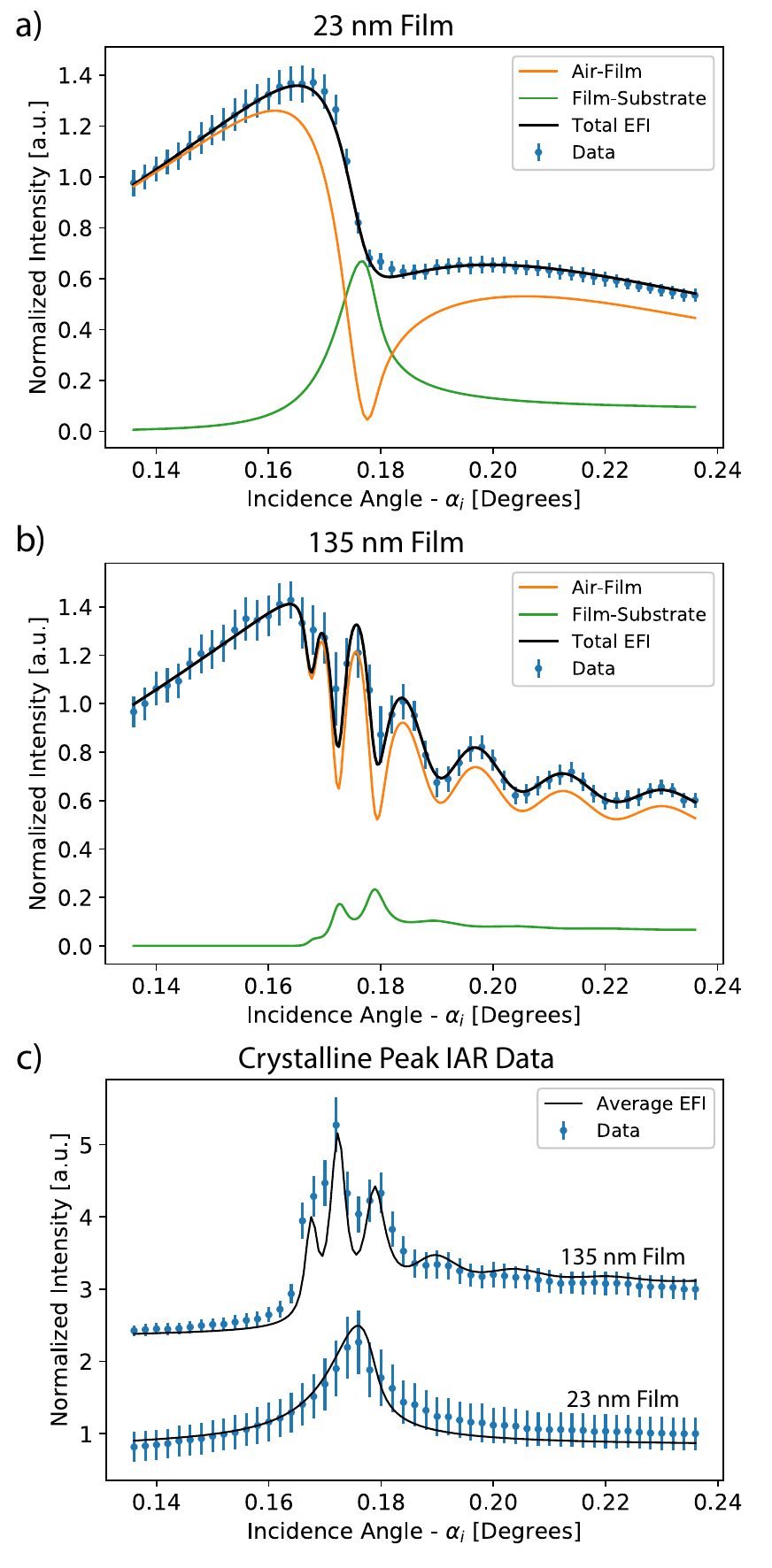}
\captionof{figure}{EFI-modulated scattering intensity data and model for a) 23 nm and b) 135 nm film thicknesses. The overall signal has contributions from the air-film and film-substrate interfaces, which is dependent on film thickness, and roughness and contrast at each interface c) For each film, the average EFI is plotted alongside IAR data generated from sampling the crystalline peak of Nafion. The agreement between data and model indicate crystallites are present throughout the film. Data in c) has been vertically shifted for clarity.}
\end{center}
\par Figure 3 shows experimental specular rod data and the fit for two films of different thicknesses, showing the contributions from the top and bottom interfaces. Nafion is a semi-crystalline polymer and in addition to the specular rod, the diffraction peak corresponding to the $\langle$100$\rangle$ lattice plane is also sampled to generate IAR data. Figure 3c plots this IAR data along with the average film EFI, using the parameters extracted from fitting the specular rod. These also show good agreement, suggesting crystallites are present throughout the entire film depth.
\begin{center}
\includegraphics[width=0.85\linewidth]{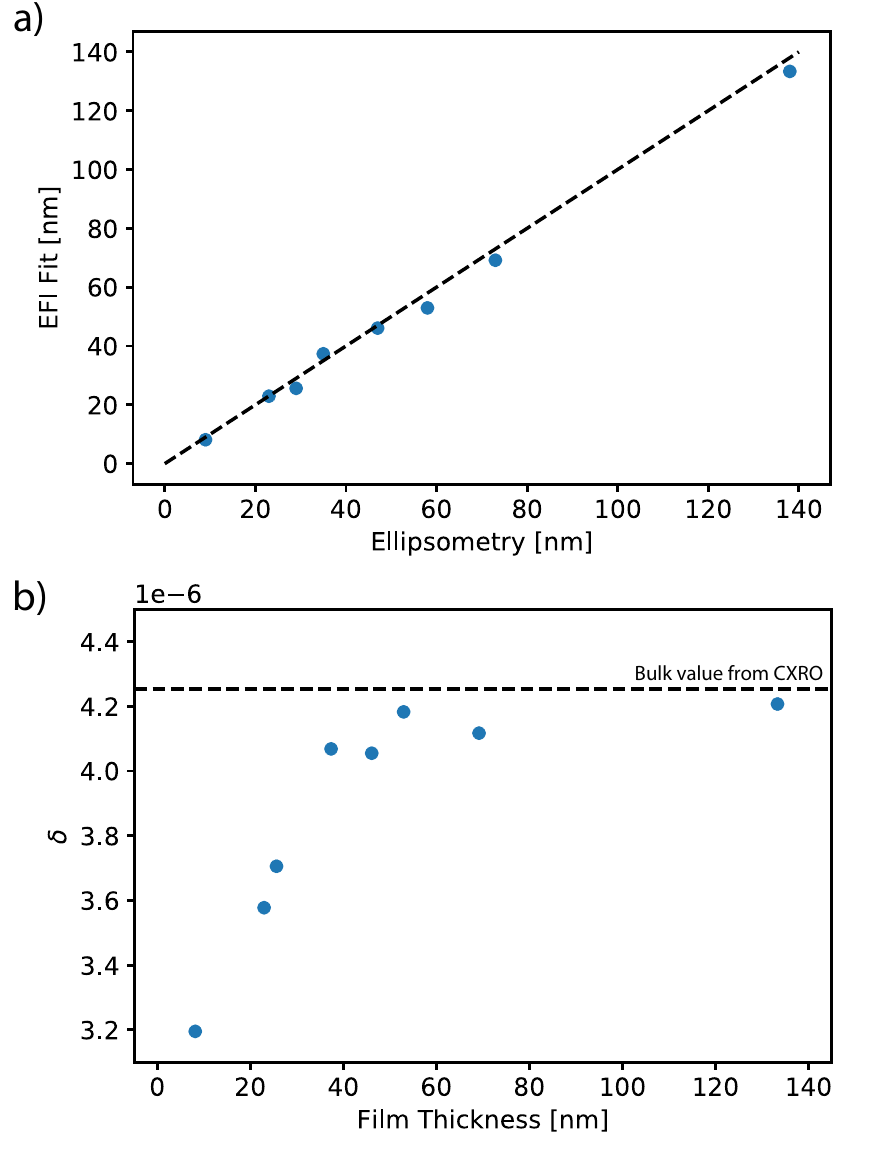}
\captionof{figure}{a) Film thickness extracted from EFI fitting plotted versus film thickness measured via ellipsometry. The dashed line is a 1:1 value between the two axes b) Real part of the index of refraction plotted versus film thickness at 10 keV. The dashed line represents the $\delta$ value estimated from CXRO using bulk Nafion density ($\rho$ = 2.1 g cm$^{-3}$).}

\end{center}
\par Each of the 8 samples' specular rod data is fit, and the film thickness from fitting is plotted versus film thickness measured via ellipsometry (Figure 4a). Across an order of magnitude in film thickness, there is excellent agreement between the two techniques. Continuing, the real part of the index of refraction at 10 keV is plotted versus film thickness in Figure 4b. At larger film thickness, $\delta$ is close to the value estimated from CXRO using the bulk membrane density of Nafion. Below 40-50 nm, $\delta$ steeply decreases as the films become less dense; this transition point in $\delta$ occurs in the same thickness range where prior studies observed changes in this material's physical properties. \cite{yager,DECALUWE}.

\subsection{Bilayer Samples}
\par Two bilayer samples were prepared by spin casting Nafion and poly(styrene-b-butadiene-b-styrene) (SBS) layers on top of each other. These materials were chosen because they both exhibit scattering peaks in the accessible q-region for 3.6m GISAXS, and can be cast from orthogonal solvents to form a macroscopically smooth thin film. Nafion is spun cast from a water/n-propanol solution and is a weakly phase separated material with a broad isotropic ring at 2.5 nm$^{-1}$. SBS is spun cast from a toluene solution and exhibits lamellar morphology with an anisotropic peak centered in plane at 0.17 nm$^{-1}$. 
\begin{center}
\includegraphics[width=\linewidth]{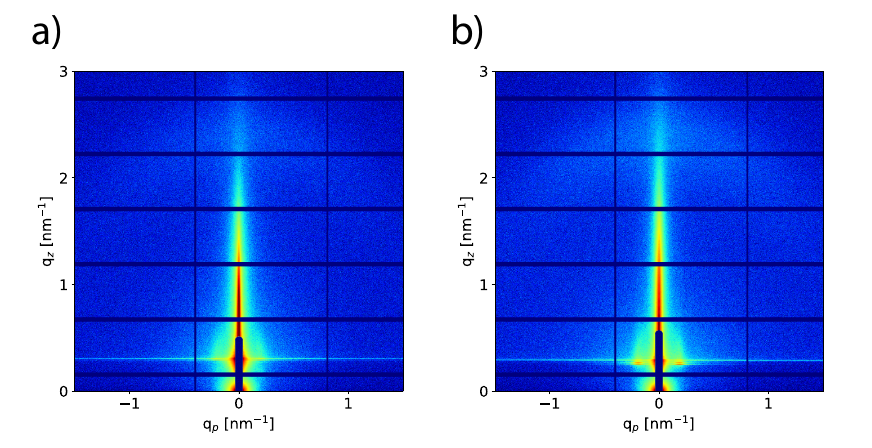}
\captionof{figure}{2D GISAXS images of Nafion-SBS bilayer thin films. a) Scattering of Sample 1 at $\alpha_i$ = 0.163$^\circ$ b) Scattering of Sample 2 at $\alpha_i$ = 0.172$^\circ$. The incidence angles were chosen such that visibility of both scattering peaks is maximized.}
\end{center}
Sample 1 is SBS spun cast first, then Nafion; Sample 2 is Nafion spun cast first, then SBS. Figure 5 shows the 2D GISAXS images for both samples. There are only small differences in the scattering patterns, and it would be difficult to discern the order of each layer from a single image. To distinguish between them, data is collected as a function of incidence angle to generate IAR data. Three regions of interest were chosen for these samples. One at the scattering peak due to Nafion, one at the scattering peak due to SBS, and one at the specular rod. Figure 6a/b shows the IAR plots for the three regions of interest.
\begin{center}
\includegraphics[width=\linewidth]{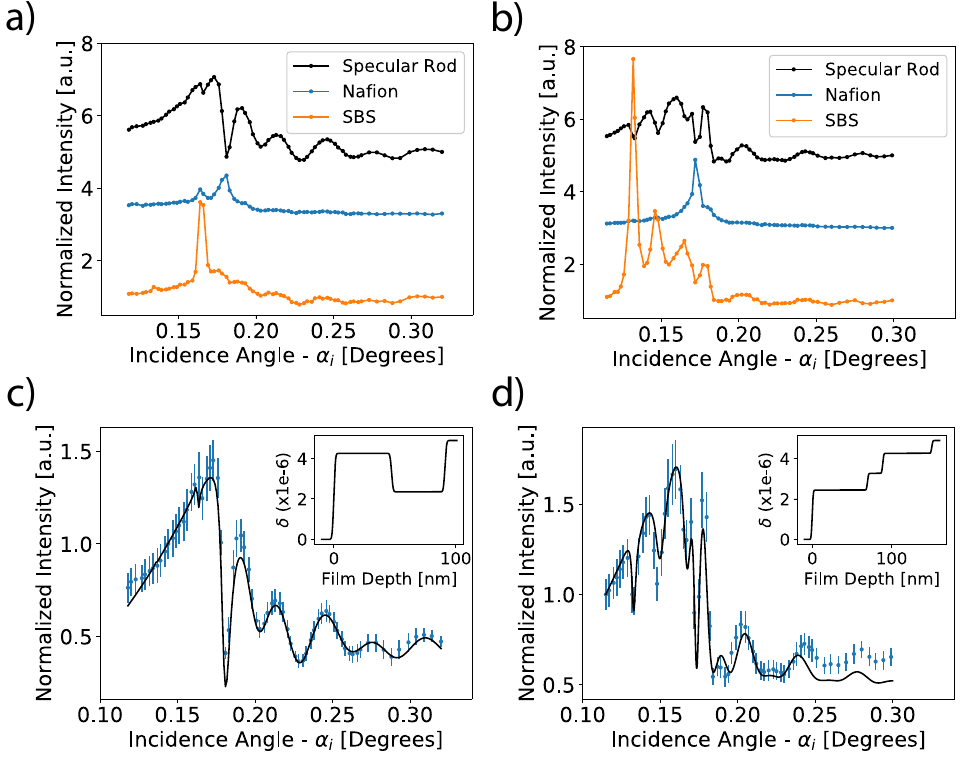}
\captionof{figure}{a), b) IAR data sets for each sample. c), d) Fits of the specular rod data with the resulting electron density profile shown inset.}
\end{center}
\par Each data set is normalized by their last value to compare the degree of signal enhancement/modulation. All three IAR data for each sample show different modulations in the scattering intensity corresponding to their unique z-positions within the sample. There is a clear difference between the specular rod IAR data for Sample 1 and Sample 2 that is less apparent with a single GISAXS image. SBS IAR data between samples is also distinctly different, highlighting that SBS is present at different depths between samples. In Sample 2, the higher electron density of the Nafion underlayer creates multiple TE modes in the SBS with an almost 8 times higher intensity at the 1st TE mode compared to higher incidence angles. For weakly scattering samples, shooting at the incidence angle for a specific TE mode could be useful for enhancing scattering above the background.

\par To extract further information, the specular rod IAR data is fit using the same differential evolution algorithm. The fitting parameters are the thickness of each layer, the index of refraction of SBS, and the relative contribution from the buried interfaces. The index of refraction for Nafion is held constant at the bulk film value because these layers are expected to be thicker than the transition point observed in the single layer films.
\par The specular rod fit and extracted electron density profile for Sample 1 are shown in Figure 6c, which is fit well with a two layer model. For Sample 2 a 20 nm-thick interlayer of intermediate electron density is needed between the SBS and Nafion layers to fit adequately the specular rod data (Figure 6d). This layer is not present in Sample 1, indicating that Nafion is partially soluble in toulene while SBS is insoluble in water/nPA. Though all of the features in the data are recreated, further agreement at $\alpha_i > 0.225^\circ$ may be achieved by also fitting Nafion's index of refraction. Contributions to the total specular rod signal from each of the interfaces for Samples 1 and 2 are plotted in Figure 7. 
\begin{center}
\includegraphics[width=0.85\linewidth]{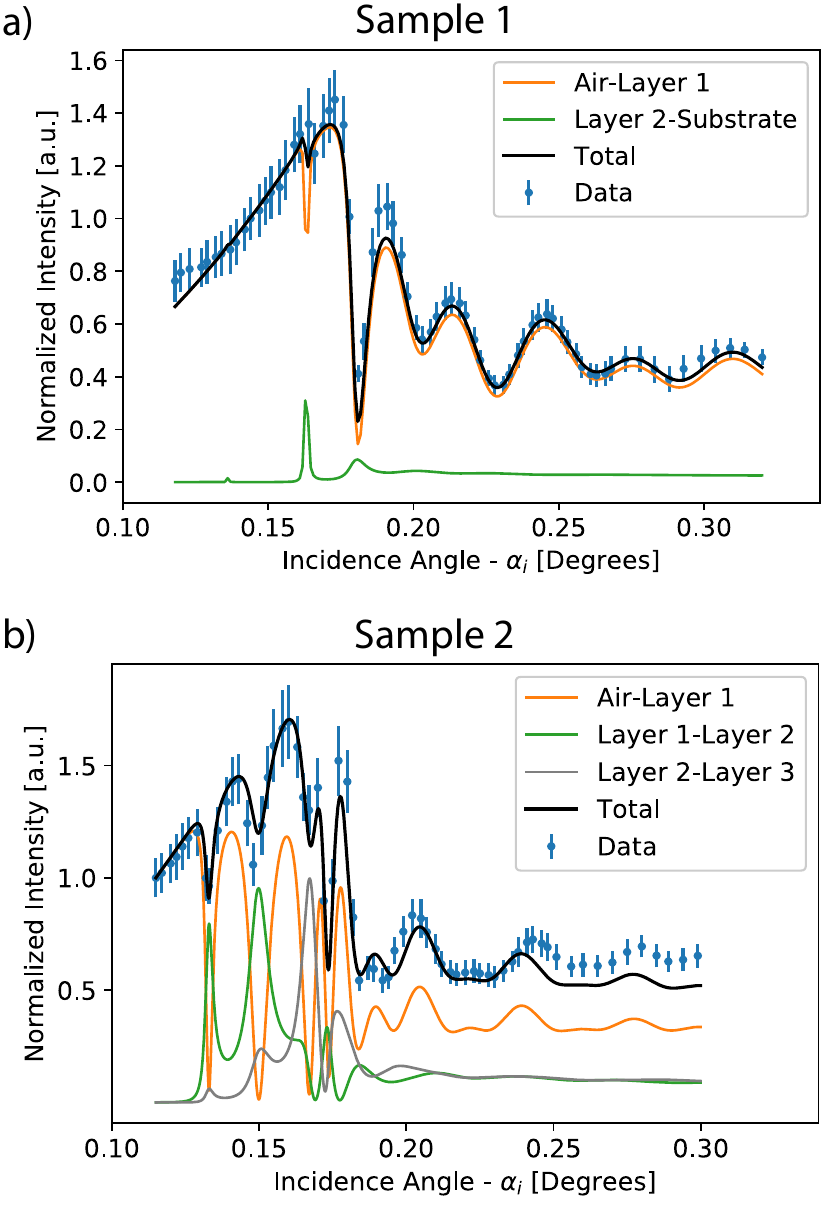}
\captionof{figure}{Breakdown of contributions to the total specular rod signal for a) Sample 1 and b) Sample 2.}
\end{center}
\par Each sample has a unique combination of EFI signal from the different interfaces which is a function of layer ordering and roughness. In Sample 1, the Nafion/SBS interface does not appear to contribute to the specular rod signal. As x-rays propagate downwards and encounter this interface, they are more likely to refract/scatter downward into the lower electron density SBS underlayer, rather than upwards which would contribute to the specular rod signal. Moreover, the bottom-layer/silicon interface has a small contribution to the signal for Sample 1 and no apparent contribution to the signal for Sample 2. This is ascribed to polished silicon's low surface roughness, especially in comparison to polymer/polymer and polymer/air interfaces.

\par Figure 8 shows a schematic of each film and the full EFI maps, which help illustrate where specific peaks in the IAR data occur spatially within the film (see Figure S3 for direct comparison of Nafion and SBS data with average layer EFI). The layering of Sample 1 leads to interesting and subtle features observed in the IAR data. The dip in the specular rod signal and peaks in the Nafion and SBS IAR data at $\alpha_i $ = 0.163$^\circ$ is unexpected (Figure 6a), since this is below the critical angle of Nafion, but is explained through the full EFI map. At this incidence angle the evanescent wave penetration depth is high enough to reach and couple into the SBS underlayer leading to TE modes in the SBS layer and higher EFI in the Nafion layer. EFI-modulated scattering could be applied to thin film waveguides to study behavior like this in more detail.
\begin{center}
\includegraphics[width=0.85\linewidth]{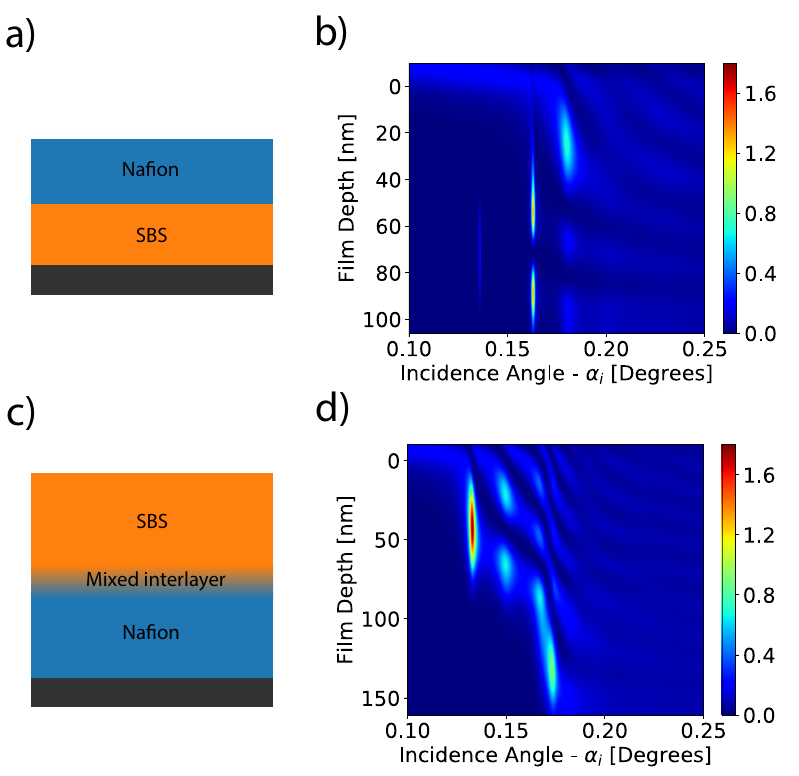}
\captionof{figure}{a), c) A schematic of each film, illustrating the layer ordering and b), d) full EFI maps for each sample. }
\end{center}
\par Using EFI-modulated scattering, the difference between these two samples is easily identified and the source of each scattering signal can be mapped depth-wise by fitting IAR data. Significantly more information is elucidated when compared to a single or few scattering images, and the data can be modeled without any knowledge of form or structure factors. The technique is inherently \textit{in situ}, and does not require any special experimental set-up beyond what is required for GIXS. 
\par However, to complete morphology studies under environmental control (e.g. temperature- or solvent vapor-annealing), additional experimental equipment is normally required to track film thickness \textit{in situ}, or the experiment must be repeated \textit{ex situ} under the same conditions. With EFI-modulated scattering, film thickness and index of refraction can be tracked \textit{in situ} to connect macroscopic properties with nanostructural morphology changes in one experiment. The intense specular rod signal allows for short exposure times such that the temporal resolution is limited by the goniometer motor speed. Transport and thermodynamic phenomena occurring on the order of seconds to minutes can be depth-resolved, including nanoparticle diffusion and the ordering of myriad polymer systems. 
\par Beyond \textit{in-situ} experiments, EFI-modulated scattering can be applied to characterize any thin film system. Organic photovoltaics, polymer nanocomposites for separations, and layer-by-layer films for biomedical applications are all examples of systems where critical processes occur through the depth of the film. EFI-modulated scattering has the ability to define through-plane ordering with considerable detail, which opens up a new paradigm for characterization and enables more robust structure-property relationships in these and other materials.

\section{Beamline Procedure}
\par This technique can be translated into a beamline plugin for on-line data analysis: A portion of the specular rod is selected on the detector over which to integrate intensity. An incidence angle scan is then performed at short exposure times, integrating the region of interest (ROI) for each angle, and returning the IAR data to the user. From these plots, critical angles and TE modes are easily identified, and the researcher can make a more informed decision on which incidence angles to collect longer exposures (e.g. at the 1st TE mode to maximize signal, or a specific TE mode to highlight some depth of the film). The current method reduces a set of 2D-data off-line; immediately returning the IAR data reduced into a text file will reduce the amount of data that needs processed by the researcher and reduce the amount of physical space required on data servers.

\par Some grazing-incidence scattering beamlines have a diode installed as a reflectivity point detector in addition to an area detector, which would make sampling the specular rod somewhat redundant. Where this technique becomes useful even on beamlines with simultaneous reflectivity is the ability to sample multiple ROIs on the area detector. Multiple scattering peaks can be sampled simultaneously to generate IAR data for each peak that could then be co-refined with reflectivity data to precisely characterize the film as a function of depth.
\par For this technique to be applied successfully, a few conditions are necessary. The first requirement is an x-ray source of small angular divergence and energy FWHM. As these increase, the magnitude of EFI signal modulation becomes smaller. The second criteria recommended is a goniometer with uncertainty in the incidence angle of less than $\sim$5\% ($\sim$0.005$^{\circ}$ for 10 keV). Uncertainty in the motor position could shift data points in a non-uniform way and lead to poorly resolved data. For longer sample-to-detector distances where the angular resolution per pixel is higher, an alternative strategy is to use the substrate’s Yoneda peak position as an internal standard for calculating the incidence angle of each image.

\section{Summary}
\par In this paper we review the multilayer DWBA theory, and use its framework to show how incidence angle-resolved (IAR) data is proportional to the local electric field intensity (EFI). Applying these principles, we demonstrate how EFI-modulated scattering can be used to extract optical constants and film thickness from Nafion polymer thin films ranging from 10 - 140 nm. Bilayer thin films of Nafion and SBS are analyzed to show how multiple sources of scattering can be depth-resolved within a sample. EFI-modulated scattering is a general technique, enabling \textit{in-situ} optical characterization and depth profiling for any thin-film system.

\subsection*{Acknowledgements} 
The authors would like to thank Zhang Jiang, Chenhui Zhu, and Alex Hexemer for helpful discussions. The authors would also like to thank Alastair MacDowell for providing the multilayer monochromator specifications. This work is supported by the Army Research Office under award number AWD00000675. This research used beamline 7.3.3 of the Advanced Light Source, which is a DOE Office of Science User Facility under contract no. DE-AC02-05CH11231.

\bibliographystyle{ieeetr}
\bibliography{main.bib}

\end{multicols}

\end{document}